\documentclass[prl,aps,twocolumn,superscriptaddress,showpacs,preprintnumbers,nofootinbib]{revtex4-1}
\usepackage{amsmath,amssymb}
\usepackage{graphicx}
\usepackage{tabularx}
\newcolumntype{Y}{>{\centering\arraybackslash}X}

\usepackage{array}   % for \newcolumntype for math mode tables
\newcolumntype{L}{>{$}l<{$}} % math-mode version of "l" column type
\newcolumntype{C}{>{$}c<{$}} % math-mode version of "c" column type
\newcolumntype{P}{>{$}p{2cm}<{$}} % math-mode version of "p" column type

\usepackage{mathrsfs}
\usepackage{dsfont}
\usepackage[pdftex,colorlinks=true,linkcolor=black,citecolor=black,urlcolor=black]{hyperref}
\usepackage{enumerate}
\usepackage{epstopdf}
\usepackage{slashed}
\usepackage[normalem]{ulem}

\def\eq#1 { \begin{equation} #1 \end{equation} }
\def\eqn#1{ \begin{align} #1 \end{align} }

\def\a{\alpha}

\def\g{\gamma}
\def\e{\epsilon}

\def\s{\sigma}

\def\d{\partial}

\def\cN{\mathcal{N}}

\def\sl2r{SL(2,\mathbb{R})}

\def\ce{\varepsilon}

\newcommand{\tr}{{\mathrm{tr}}}

 \newcommand{\be}{\begin{equation}}
\newcommand{\ee}{\end{equation}}

\usepackage{bm}

\usepackage{mathtools,cancel}
\usepackage[usenames,dvipsnames,svgnames,table]{xcolor}

\newcommand{\mc}{\mathcal}

\usepackage{marvosym}
\usepackage{tikz}
\usepackage{tikzsymbols}
\usepackage{phaistos}
\makeatletter
\def\@fnsymbol#1{\ensuremath{\ifcase#1\or \,\or \text{\Cat}  \or
   \mathsection\or \mathparagraph\or \|\or **\or \dagger\dagger
   \or \ddagger\ddagger \else\@ctrerr\fi}}
\makeatother
\renewcommand{\thefootnote}{\fnsymbol{footnote}}
\newcommand{\bgs}{{\sigma_{\mathrm{bg}}}}

\newcommand{\bgD}{{D_{\mathrm{bg}}}}
\newcommand{\bgF}{{F_{\mathrm{bg}}}}
\newcommand{\bgp}{{\phi_{\mathrm{bg}}}}
\begin{document}

%%% Sets the spacing above and below equations %%%
\setlength{\abovedisplayskip}{9pt}
\setlength{\belowdisplayskip}{9pt}
%%%%%%%%%%%%%%%%%%%%%%%%%%%%%%%%%%%%%%%%%%%%%%%%%%

\preprint{Imperial/TP/2019/MMR/01}
\title{Supersymmetric space-time symmetry breaking sources} 

\author{Louise Anderson}
\affiliation{Stanford Institute of Theoretical Physics, Stanford University, Stanford, CA 94305, USA}
\author{Matthew M.~Roberts}
\affiliation{Blackett Laboratory, Imperial College, London, SW7 2AZ, U.K.}

\date{\today}

%%%% restarts footnote counter to normal numbers %%%%%%%%
\renewcommand{\thefootnote}{\arabic{footnote}}
\setcounter{footnote}{0}

\begin{abstract}

We construct  new families of deformed supersymmetric field theories which break space-time symmetries but preserve half of the original supersymmetry. We do this by writing deformations as couplings to background multiplets. In many cases it is important to use the off-shell representation as auxiliary fields of the non-dynamical fields must be turned on to preserve supersymmetry. We also consider backgrounds which preserve some superconformal symmetry, finding scale-invariant field profiles, as well as $\mathcal{N} =2$ theories on $S^3$. We discuss how this is related to previous work on interface SCFTs and other holographic calculations.

\end{abstract}

\maketitle

%%%%%%%%%%%%%%%%%%%%%%%%%%%%%%%%%%%%%%%%%%%%%%%%%%%%%%%%%%%%%%%%%
{\em Introduction}--- 
%%%%%%%%%%%%%%%%%%%%%%%%%%%%%%%%%%%%%%%%%%%%%%%%%%%%%%%%%%%%%%%%%
An open question in theoretical physics is the classification of renormalisation group (RG) flows when turning on deformations, or source terms, which break space-time symmetries. Much of what we know about RG flows is restricted to the case of sources which preserve Lorentz invariance, and much less is known in cases when we break it, owing to the lack of a momentum space description in inhomogeneous systems. One may hope to gain analytic traction by studying deformations which preserve supersymmetry. It is well known that many coupling constants in supersymmetry QFTs can be interpreted as non-dynamical background supermultiplets, and this can be used to prove powerful non-renormalisation theorems for Lorentz invariant systems \cite{Seiberg:1993vc}. For example, in \cite{Babington:2005vu}, the authors studied $\cN=1$  with space-dependent gauge- and superpotential coupling arising from background chiral multiplets.  Similarly, spatially dependent couplings were also studied in  \cite{Kraus:2001tg, Kraus:2001id, Kraus:2002nu, Kraus:2002qq, Kraus:2002se, Kraus:2001kn} where they were shown to  give rise to an additional new anomaly at one loop. 

Supersymmetric theories with space-dependent couplings also naturally arise in the study of superconformal field theories with defects and/or boundaries \cite{Erdmenger:2002ex}. This leads to a rich zoo of so-called Janus solutions, where the couplings take different values on each side of a domain wall. These have been considered in the context of 4d $\cN>2$ super-Yang-Mills theory in \cite{Clark:2004sb,DHoker:2006qeo,Kim:2008dj,Gaiotto:2008sd,Choi:2017kxf}. 
In the special case of the maximally supersymmetric ABJM theory, there  exist a class of spatially dependent mass-deformations \cite{Kim:2018qle, Kim:2019kns}, dual to the supersymmetry Q deformations of \cite{Gauntlett:2018vhk, Arav:2018njv}, which preserve at most half of the original super-Poincar\'e symmetry.  However, the derivation of these deformations are often complicated and done through a brute force method. It is therefore important to ask if one could take a systematic approach to these using coupling to spatially dependent background mutliplets, and thereby classify more general deformations.

Another interesting setting in which the role of  spatially dependent couplings  have been  less investigated is in the context of localisation in Euclidean 3d $\cN =2$ theories. These theories have been coupled to background supergravity multiplets  to obtain supersymmetric theories on curved spaces in the rigid limit  \cite{Festuccia:2011ws}. 

In this work, we will use this approach of coupling to background multiplets to find a new class of spatially modulated deformations of  3d $\mc{N}=2$  theories preserving half of the original supersymmetry, expanding the results of \cite{Kim:2018qle} to more general theories. We focus on Lorentzian theories, but also consider Euclidean $\cN=2$ theories on $S^3$ and demonstrate that a similar construction breaking the space-time $SO(4)$ but preserving some supersymmetry is possible.  We also consider 4d, $\mc{N}=1$ theories. We find that this analysis reproduces known interface configurations as well as new, nontrivial, deformations which, as in the 3d case, break space-time symmetries but preserve half of the original supersymmetry. We also study when such deformations can preserve any superconformal symmetry, yielding scale-invariant interfaces and defects. We then conclude with a discussion and comment on future directions.

%%%%%%%%%%%%%%%%%%%%%%%%%%%%%%%%%%%%%%%%%%%%%%%%%%%%%%%%%%%%%%%%%
{\em Background fields and supersymmetry}---
%%%%%%%%%%%%%%%%%%%%%%%%%%%%%%%%%%%%%%%%%%%%%%%%%%%%%%%%%%%%%%%%%
Coupling a gauge theory to background fields to introduce dimensionful parameters to the theory is a long-standing story, as described in the introduction. The classic approach to this is to demand that these background fields are invariant under the entire supersymmetry algebra,  and as such are Lorentz invariant, but one could demand less. Herein, we will instead only  ask that these background fields are invariant under a subset of the original transformations, resulting in a less supersymmetric theory.  A well known example of this is the case of turning on masses in a superconformal field theory, where the mass term can be viewed as turning on a background field breaking conformal invariance but preserving the Poincar\'e generators. 

One could also approach situations where these background fields are spatially modulated along the spirit of \cite{Festuccia:2011ws}. 
It is quite surprising that we can preserve half of the supersymmetry of the original theory while drastically reducing the preserved space-time symmetries. Under most circumstances, it is even possible to introduce spatially modulated deformations that remain invariant under some subset of  superconformal transformations. 

The benefit of rewriting couplings in the Lagrangian in terms of background multiplets is that it makes determining the amount of remaining supersymmetry quite straight-forward. One simply takes the supersymmetry transformation rules for the full interacting theory, and any such transformation under which the background multiplet is invariant will be a bona-fide supersymmetry of the dynamical theory. As we can only turn on classical expectation values for bosonic fields, this reduces to looking at the variations of the fermions, also referred to as BPS equations. Note that this sometimes requires  modifications of transformations of dynamical matter fields from making derivatives covariant. For example in a dynamical matter multiplet in 3d $\cN=2$ theories,  \eq{
\delta_\e \psi \supset - ( \g^\mu\nabla_\mu \phi )\e \rightarrow -  \g^\mu (\nabla_\mu + i A^\mathrm{bg}_\mu) \phi - \bgs \phi .}

Strictly speaking this analysis is semi-classical, and there can be quantum corrections from loops, for instance as dynamical fields get anomalous dimensions. However unless we have a case with spontaneous supersymmetry breaking, such effects will, as expected, just cause the space-dependent couplings to be renormalised as we flow to the infrared. One can consider the construction in this paper as classifying the possible spatially modulated UV couplings. It is worth pointing out that we could, even in cases where the theories do not admit a Lagrangian description, consider coupling a background vector multiplet to conserved currents from global symmetries in the vein of \cite{Cordova:2016xhm}.

Since we here are interested in supersymmetric deformations breaking space-time symmetries, we will consider subsets of the full supersymmetry algebra defined by projectors of the Clifford algebra with preferred orientation in space-time, for example $v_\mu \g^\mu \e  \propto \e$ with a preferred vector $v^\mu$ breaking Lorentz invariance. This presents us with three obvious cases: $v$ can be time-like, space-like or null. However, we will see that the Clifford algebra will often make some of these cases redundant. 

Superconformal transformations are instead generated by conformal Killing spinors, which in flat space, in dimensions greater than two, take the form $\e_S =x_\mu  \g^{\mu} \chi$ for some constant spinor $\chi$. By considering variations by such spinors, we are able to make statements about the residual superconformal structure for certain scale invariant background fields. This allows us to  reproduce standard Janus-type results, as well as finding a much more general class of scenarios. Our results can be easily adapted to Euclidean signature, even on some curved backgrounds, such as $S^3$.

%%%%%%%%%%%%%%%%%%%%%%%%%%%%%%%%%%%%%%%%%%%%%%%%%%%%%%%%%%%%%%%%%
{\em  $\mathcal{N}=2$ theories in 2+1 dimensions}---
%%%%%%%%%%%%%%%%%%%%%%%%%%%%%%%%%%%%%%%%%%%%%%%%%%%%%%%%%%%%%%%%%
Background fields in three dimensions are naturally used as mass terms for dynamical fields.  As in \cite{Aharony:1997bx, Gomis:2008vc} we can write $\cN=2$ mass terms either as D-term or F-term, which can be thought of as arising from background vector or matter multiplets respectively. If we turn on the scalars $\bgs,~\bgD$ of a background vector for a $U(1)$ that matter fields are charged under, we induce mass terms via the interactions 
\eq{
\label{eq:3d_interaction}
\mathcal{L}_{int}=  
- \bar{\chi} \bgs \chi 
- \bar{\phi}\left(\s_\mathrm{bg}^2-\bgD \right) \phi
+ \text{h.c}
} where $\chi, \phi$ are the dynamical chiral multiplet fermions and scalars respectively and we have suppressed traces over gauge and flavour indices. Similarly if we write a superpotential mass term as a coupling of dynamical multiplets to a background matter multiplet $M$ with scalars $\bgp,~\bgF$, we get
\eq{
\int d^2\theta M \Phi^2 \supset \bgp (\psi^2 +\phi F) + \bgF \phi^2 
}
We can of course consider other couplings, such as the  classically marginal $\lambda \Phi^4$, and will keep the dimension of the fields general when considering superconformal transformations.

In these three dimensional theories the supersymmetry transformations are parametrised by a complex Dirac spinor. Again, we have three possibilities for the preferred space-time direction we choose for our projection condition: time-like, null or space-like. However, it turns out that the Clifford algebra in $2+1$ dimensions gives  
\eq{
\g^2 \e = \pm \e \quad \Leftrightarrow \quad
(\g^0 \pm \g^1)\e= 0,\label{eq:3d_spacelike_proj}
}
so the space-like and null projectors are equivalent. Furthermore, the time-like condition is equivalent to a complex projector,
\eq{
\g^0 \e = \pm i \e 
\quad \Leftrightarrow \quad 
( \g^1 \pm i \g^2) \e  = 0. \label{eq:3d_timelike_proj}
}
We can also write projection conditions using charge conjugation conditions, such as $\epsilon = \pm \epsilon^*$, $\gamma^2 \epsilon^* = \pm \epsilon$ at the cost of breaking the $U(1)_R$ symmetry. After chosing a projector we will then find the most general backgrounds which preserve half of the original $\cN=2$  by requiring the vanishing of the BPS equations.

For an $\mc{N}=2$ background vector multiplet, the supersymmetry variation of the gaugino is proportional to
\eq{\label{eq:3d_vec_vary}
\frac{1}{2} \ce_{\mu \nu \rho} \gamma^{\rho} \epsilon \bgF^{\mu \nu}   +  \epsilon \bgD +  \gamma^\mu \epsilon  \nabla_\mu \bgs   + \frac{2}{3} \gamma^\mu (\nabla_\mu \epsilon) \, \bgs
.}
For simplicity, we will only consider turning on background fields for a a $U(1)$ factor of the global symmetry and turn off the gauge field to focus on mass terms. 

The time-like projector  leads to no non-trivial solutions, as do projectors which break $U(1)_R$. The space-like projector gives 
\eq{
\g^2 \e_\pm = \pm \e_\pm \rightarrow \bgs = \bgs(t \mp x, y),~ \bgD = \mp \d_y \bgs.
,}
This allows for a wide class of spatially modulated masses. The result is strikingly similar to the supergravity result of \cite{Gran:2008vx}, which could be describing a bulk dual to this deformed theory. Note that even if we restrict the background to invariant under $ISO(1,1)$ we do not preserve any more supersymmetry, but of course in theories with larger $\cN$ this will not be the case. For instance in the ABJM theory with a spatially varying mass deformation \cite{Kim:2018qle} preserving half the super-Poincar\'e symmetry requires either $m(y)$ or $m(t\mp x)$, but not $m(t \mp x,y)$ as we have here.

If we want the background to have some residual scale invariance, we need some variations by $\e_S = x_\mu \g^\mu \chi$ for a constant spinor $\chi$ to vanish. Consider the case with preserved super-Poincar\'e $\g^2 \e_Q = +\e_Q$. We find space-like or null singular sources
\eqn{
\g^2 \chi = -\chi  & \quad \rightarrow\quad \bgs = \frac{\lambda}{y},~\bgD = \frac{\lambda}{y^2}, \\  
\g^2 \chi_= +\chi &\quad\rightarrow\quad \bgs = \frac{\lambda}{t-x},~\bgD = 0.
}
Note the space-like case matches the boundary analysis of M-theory Janus solutions described in \cite{Arav:2018njv}.

Now  consider background chiral multiplets. Assuming it is not charged under any global symmetries with background vectors turned on (including this requires making the derivatives covariant, we ignore it for simplicity) the fermion variation is 
\eq{ -\gamma^\mu \nabla_\mu \bgp \epsilon - \frac{2\Delta}{3}  \gamma^\mu \left( \nabla_\mu \epsilon \right) \bgp +\bgF \epsilon^*}
where we have allowed for a non-canonical scaling dimension $\Delta$. The analysis is similar to the vector case, only now our scalar fields are complex. We have the time-like projector
\eqn{
\g^0 \e = \pm i \e & \quad \rightarrow\quad \bgp=\bgp(x\pm i y),~\bgF=0, \label{eq:3d_chiral_cx} 
}
which when we require some superconformal symmetry gives
\eqn{
\g^0 \chi = \pm i \chi&  \quad \rightarrow\quad \bgp= \frac{\lambda}{(x\pm i y)^\Delta}.
}
We can also consider the space-like projector and find
\eqn{
\g^2 \e = \pm \e &  \quad \rightarrow\quad  \bgp=\bgp(t\mp x),~\bgF=0, 
\\  
\g^2 \chi = \pm \chi &  \quad \rightarrow\quad  \bgp = \frac{\lambda}{(t\mp x)^\Delta}.
}
We can also consider projection conditions which break $U(1)_R$, and the only non-trivial case is
\eqn{
\g^2 \e^* = \pm \e &  \; \rightarrow\; \bgp = \bgp(y),~ \bgF = \pm \bgp'(y), 
\\  
\g^2 \chi^* = \mp \chi &  \;\rightarrow\; \bgp = \frac{\lambda}{y^\Delta},~ \bgF = \mp \frac{\Delta \lambda}{y^{\Delta+1}}.
\label{eq:chiral_3d_full_ISO(1,1)}
}
Note that unlike the vector case, in \eqref{eq:chiral_3d_full_ISO(1,1)}, the preserved supersymmetry algebra includes the full $ISO(1,1)$ of the $(t,x)$ plane at the cost of breaking the $R$-symmetry. We also have one projection condition, \eqref{eq:3d_chiral_cx}, whose background has no null isometries. From a holographic dual point of view this is a rather surprising scenario as the analysis of \cite{Gran:2008vx} indicates that supersymmetric gauged supergravity solutions which we might interpret as bulk duals with spatially varying sources turn on always have a null killing vector, but we see no sign of it here. We leave a holographic comparison of this case to future work. One could consider turning on background gauge fields as well, however, this does not alter the spirit of our results. It often amounts to requiring self-dual background gauge fields with the same space-time isometries as the scalars.  
 
%%%%%%%%%%%%%%%%%%%%%%%%%%%%%%%%%%%%%%%%%%%%%%%%%%%%%%%%%%%%%%%%%%
{\em  $\cN =2 $ theories on $S^3$}---
%%%%%%%%%%%%%%%%%%%%%%%%%%%%%%%%%%%%%%%%%%%%%%%%%%%%%%%%%%%%%%%%%%
Another very interesting question is if this analysis could be reproduces on other backgrounds than flat space. Of special interest is $S^3$, since this immediately hints at the possibility of using supersymmetric localisation to compute observables in these deformed theories exactly \cite{Pestun:2007rz}.

The Euclidean construction of $\cN=2$ theories  is slightly different \cite{Kapustin:2009kz}, as due to the lack of a Majorana condition in 3+0 dimensions, there is a doubling of degrees of freedom, for a review see \cite{Hama:2010av,Marino:2011nm} and a holographic example in \cite{Freedman:2013ryh}. For a vector multiplet, this means that $\phi$ and $D$ are complex, and there are two independent complex supersymmetry parameters $\e,~\bar\e.$  The superconformal algebra in flat Euclidean space is replaced with the symmetry generated by conformal Killing spinors on $S^3$, and the full set of possible supersymmetry parameters are the left- and right-invariant spinors, and the conformal Killing spinors which are constant when working in the left- or right-invariant frame respectively. Let us again consider $D$-term masses arising from a background vector multiplet. The BPS equations are, setting $\bgF=0$,
\eqn{
\left(- \bgD  + i \g^\mu \nabla_\mu \bgs + \frac{2i}{3}\bgs \g^\mu \nabla_\mu \right)\varepsilon=0,
}
for both variations $\varepsilon=\e,\bar\e$. For constant background fields, as opposed to Minkowski space,  we can only preserve half of the $SU(2)_L \times SU(2)_R$, as we find
\eq{
\e_L,\bar\e_L:~ \bgs + \bgD = 0,~ \e_R,\bar\e_R:~ \bgs- \bgD=0.
}
As in our Lorentzian analysis let us consider a projection condition, e.g. $\g^3 \e_L = + \e_L$ in the left-invariant frame  $e^\mu_i$. The BPS equations reduce to
\eq{
(e_1 + i e_2)^\mu \d_\mu \s = 0,~ \bgs - i e_3^\mu \d_\mu \bgs+\bgD= 0.
}
Thanks to the symmetry of the three-sphere we can solve this PDE exactly. In standard Hopf coordinates $ds^2 = d\eta^2 + \sin^2\eta d\xi_1^2 + \cos^2\eta d \xi_2^2$, this has regular solutions
\eq{
\bgs =\sum_{m>0,n>0} \s_{m,n} e^{i m \xi_1 - i n \xi_2} \sin^m \eta \cos^n \eta,
}
where positivity of $m,n$ is required for the function to be smooth at the poles. The residual supersymmetry parameter squares to the complex Killing vector $(e_1 + i e_2)^\mu$. 
Therefore, there are spatially dependent  mass deformations that are 1/4 BPS on $S^3$. An analogous analysis can be done for $F$-term masses coming from a background matter multiplet, and the results are similar. 

In a theory with only $\cN=2$, these kinds of  deformations would however not leave enough supersymmetry for a localisation calculation to be carried out. It would therefore be very interesting to consider how these 1/4 BPS deformations behave in theories with  $\cN>2$ supersymmetry. The above results indicates that one could preserve enough supersymmetry to allow for localisation calculations to be carried out in theories with more supersymmetry. It would also be interesting to construct holographic duals to theories with such deformations, and see if these kinds of deformations could be turned on along the lines of \cite{Toldo:2017qsh}.

%%%%%%%%%%%%%%%%%%%%%%%%%%%%%%%%%%%%%%%%%%%%%%%%%%%%%%%%%%%%%%%%%
{\em  $\mathcal{N}=1$ theories in 3+1 dimensions}---
%%%%%%%%%%%%%%%%%%%%%%%%%%%%%%%%%%%%%%%%%%%%%%%%%%%%%%%%%%%%%%%%%
In four dimensions, background chiral multiplets can be used to describe coupling constants such as masses, Yukawa couplings, and Yang-Mills gauge couplings via  superpotential interactions, e.g. $W \supset \Phi_\mathrm{bg} \Phi^2,~\Phi_\mathrm{bg}\Phi^3,~\Phi_\mathrm{bg} \tr \mc{W}^2$. When the auxiliary background field in the supermultipet  is turned on it will act as a source term for another operator in the same multiplet, e.g. the gaugino bilinear \cite{Clark:2005te},
\eq{
\int d^2\theta \Phi_\mathrm{bg} \tr \mc{W}^2 +\mathrm{h.c.} = L_{SYM}(\tau = \bgp) + \bgF \tr \lambda^2,
}
The parameter for minimal supersymmetry is now a Majorana spinor, or equivalently, a Weyl spinor. We will here use the notation of \cite{Freedman:2012zz}, $\e_{Maj.} = (\psi_L,\psi_R)^T,$ where in the Weyl basis, $\e_R = i \sigma^2 \e_L^*$. We can again consider superconformal transformations by the conformal Majorana Killing spinors $\e_S = x_\mu \g^\mu \chi$, with $\chi$ a constant Majorana spinor, to find backgrounds with residual superconformal symmetry. 

Using the Weyl basis, the natural projection conditions are $v_\mu \bar\sigma^\mu \e_L = \e_R$. There is no solution to the time-like projector, and so we only have space-like and null cases. The chiral fermion variation is proportional to
\eq{
P_L \left( \slashed\nabla \bgp + \frac{\Delta}{2} \bgp \slashed\nabla + \bgF \right)\e.
}
In the space-like case, allowing for a general axial phase rotation $\alpha$  we find
\eqn{
\bar\sigma^3 \e_L = +e^{i\alpha} \e_R & \rightarrow \bgp=\bgp(z),~\bgF +e^{-i\a} \bgp'(z)=0, \nonumber \\
\bar\sigma^3 \chi_L =-e^{i\a} \chi_R & \rightarrow \bgp = \frac{\lambda}{z^\Delta},~ \bgF = \frac{\Delta \lambda}{z^{\Delta+1}}, \label{eq:4d_n=1_spacelike}
}
and in the null case
\eqn{
\nonumber
(\bar\s^0 - \bar\s^3)\e_L = 0 &\rightarrow \bgp = \bgp(t - z,x + i y),~\bgF=0, \\
(\bar\s^0 \pm  \bar\s^3)\chi_L = 0& \rightarrow \bgp =
\left\{\begin{array}{cc} +: & \frac{\lambda}{(x+i y)^\Delta} \\ - : & \frac{\lambda}{(t-z)^\Delta}\end{array}
\right. .
 \label{eq:4d_n=1_null}
}

Note that the first case \eqref{eq:4d_n=1_spacelike} is the field theory dual of the original supersymmetric Janus \cite{Clark:2005te} but the second one, {\eqref{eq:4d_n=1_null} and allows for couplings with nontrivial branch points on one spatial plane, as well as dependence on a transverse null coordinate. Similar deformations were noted in \cite{Choi:2017kxf} in the case of the $\cN=4$ theory.   One can follow a similar calculation for 4d $\cN=2$ vector multiplets, and find similar structures, where the scalar is restricted to be of the form $\bgp(z),~\bgD\sim \phi_\mathrm{bg}'(z)$ or $\bgp(t-z,x+iy),~\bgD=0$ This is very reminiscent of  the 4d/2d structure discovered in \cite{Beem:2013sza}, and further investigation into this is warranted.

%%%%%%%%%%%%%%%%%%%%%%%%%%%%%%%%%%%%%%%%%%%%%%%%%%%%%%%%%%%%%%%%%
{\em Conclusions and future directions}---
%%%%%%%%%%%%%%%%%%%%%%%%%%%%%%%%%%%%%%%%%%%%%%%%%%%%%%%%%%%%%%%%%
In this paper, we have discovered new classes of deformations of supersymmetric field theories in 3d and 4d which explicitly break some space-time symmetries, as well as presented a unified framework connecting these new deformations to previous more specific results. 

This work opens up for many interesting future direction for studies. For example, a very concrete question is how the result of \eqref{eq:3d_chiral_cx}  should be reconciled with the holographic results of \cite{Gran:2008vx}. Another interesting line of study is how the results of {\eqref{eq:4d_n=1_null} can be further understood, especially in relation to the results of the 4d/2d structure discovered in \cite{Beem:2013sza}.

Furthermore, it would be very interesting to see this technique  extended to cases with more supersymmetry. Extending our analysis to such theories will allow us to make connections to various well-studied brane and string theoretic systems. More supersymmetry to start with gives us more room to break supersymmetry and still have analytic control. It would also be helpful to have a brane picture of these new defect CFTs.

This would also be of special interest in the case of $\cN>2$ theories on $S^3$, since this could allow for localisation calculations to be carried out, and thereby for obtaining exact results directly in the field theory  that could be compared to calculations in holographically dual theories, enabling tests of the holographic principle in  a new class of theories where Lorentz invariance is broken. It would also be interesting in the context of $F$-theorems as the partition function on $S^3$ is conjectured to be related to the $F$ coefficient, and  having broken Lorentz invariance we expect to find non-monotonic flows. Investigating sphere partition functions in inhomogeneously deformed QFTs will shed some light on classifying non-Lorentz invariant renormalisation flows and how, if at all, we can compare Euclidean and Lorentzian systems after breaking space-time symmetries.

%{\em Acknowledgement}--- acknowledgements dont get their own section heading under PRL, weirdly...
We would like to thank Igal Arav, Mark Bugden, Jerome Gauntlett, Amihay Hanany, Victor Lekeu, Christopher Rosen, and Jonathan Source for helpful discussions. LA is supported by the Knut and Alice Wallenberg foundation. MMR is supported by the European Research Council under the European Union’s Seventh Framework Programme (FP7/2007-2013), ERC Grant agreement ADG 339140. The authors would like to thank the hospitality of the Kavli Institute for Theoretical Physics and support from National Science Foundation under Grant No. NSF PHY-1748958.

%%%%%%%%%%%%%%%%%%%%%%%%%%
\addcontentsline{toc}{section}{Bibliography}
\bibliographystyle{JHEP}
\bibliography{refs.bib}

\end{document}